# Stars throw their weight in old galaxies

*Nate Bastian – Excellence Cluster Universe, Technische Universität München*

**Nature News and Views, April 25th, 2012**

**The observation that old, massive galaxies have a larger fraction of low-mass stars than their younger, lower-mass counterparts adds to mounting evidence that star formation may have been different in the early Universe.**

When a star is born, it can have a mass 0.1 to 100 times that of the Sun. This property controls a star's influence on its environment, its lifetime and even its ability to host habitable planets. The relative distribution of low- to high-mass stars in a newly formed population, known as the initial mass function (IMF), is crucial to determining the evolution of any cosmic structure made up of stars, from stellar clusters to galaxies. In the local Universe, where individual stars can simply be counted, the form of the IMF does not seem to vary from one environment to another[1]. But it is possible that in the early Universe the IMF had a different shape. On page 485 of this issue, Cappellari et al.[2] present evidence for a different IMF in some of the oldest known objects — elliptical and lenticular galaxies. If true, this finding would have a drastic effect on our understanding of the Universe, from the formation and evolution of entire galaxies to the complex processes that shape the way that individual stars form.

Although we cannot normally observe individual stars in distant galaxies, the form of the IMF influences nearly every observable property of a galaxy. In most studies, the IMF is assumed to be universal — that is, the same as seen locally[3,4] — so that observable properties can be translated into physical ones. However, one can turn the problem around and ask what the properties of galaxies can tell us about the shape of the IMF. The light emitted by a galaxy is dominated by its high-mass stars, whereas a galaxy's mass is dominated by the much more numerous low-mass stars. Hence, if the mass and light can be measured independently, this can provide a strong constraint on the form of the IMF.

This is precisely what Cappellari and colleagues have done. They used advanced instrumentation — specifically, integral-field spectroscopy — to measure the rotational and random motions of stars in elliptical and lenticular galaxies, as a function of their position in the galaxy (a lenticular galaxy is intermediate between an elliptical and a spiral one). The stars' motion is controlled by the galaxy's gravitational potential, which is related to the galaxy's mass. By comparing the derived mass with the total light emitted by the galaxy — the mass-to-light ratio — and accounting for the gravitational effects of dark matter, the authors showed that this ratio deviates systematically from that which would be expected if the IMF were universal.

The authors' results suggest that old, massive elliptical galaxies (Fig. 1a) have a larger fraction of low-mass stars than do younger, less-massive lenticular or spiral galaxies such as the Milky Way. Low-mass stars, although faint, account for the majority of stars within galaxies, and are the most likely potential hosts of habitable planets. Although this deviation had recently been observed using independent techniques[5], it took the power of integral-field spectroscopy to eliminate the possibility that the deviations were caused by dark matter.

This is a surprising result, in part because a study[6] that used a similar technique on a sample of equally massive galaxies in the young Universe found that they had the same IMF as seen locally. However, Cappellari and colleagues' results[2] agree with a variety of other observations[5,7], adding to a growing chorus of claims that star formation may have been different in the early Universe.

The cause of this over-abundance of low- mass stars is unknown, especially given that studies of nearby stellar populations, in which stars can be easily identified, have not detected variations in the IMF (Fig. 1b). It is also difficult to understand why the IMF would be dependent on global galactic properties such as galaxy mass, as the authors suggest, when it does not seem to depend strongly on local ones[1], including stellar density and metal content.

If confirmed, these results would have far-reaching implications. Using the observed light of galaxies, and adopting some form of the IMF, researchers can estimate the rate at which stars are being born in distant galaxies, along with the total mass of all the stars in them. If the IMF is varying systematically as a function of environment, the conversion of observable properties to physical ones becomes much more complex. Uncertainty in the IMF is one of the main obstacles limiting our understanding of the evolution of galaxies and of the time during which most of the Universe's stars were formed.

However, a systematic variation in the IMF may provide us with our best chance to understand the details of the complex process of star formation[8]. Current numerical and theoretical models of star formation can reproduce the form of the IMF observed locally. One way to distinguish between this myriad of models is to see whether any can reproduce the types of variation reported by Cappellari and colleagues.

There is a long tradition of studies claiming that the IMF varies as a function of environment, especially in distant galaxies where individual stars cannot be identified directly and secondary tracers must be used. Most of these claims have fallen by the wayside when newer and better data became available, or as our understanding of the intricacies of the tracers has improved. However, the observations and analysis of Cappellari and coleagues[2] present a promising avenue that may lead to fundamental changes in the way that we understand how stars form and galaxies evolve.

1. Bastian, N., Covey, K.R. & Meyer, M.R. Annu. Rev. Astron. Astrophys. 48, 339–389 (2010).
2. Cappellari, M. et al. Nature 484, 485–488 (2012).
3. Kroupa, P. Science 295, 82–91(2002).
4. Chabrier, G. Publ. Astron. Soc. Pacif. 115, 763–795 (2003).
5. Auger, M. et al. Astrophys. J. 721, L163–L167 (2010).
6. Cappellari, M. et al. Astrophys. J. 704, L34–L39 (2009).
7. van Dokkum, P. G. & Conroy, C. Nature 468, 940–942 (2010).
8. McKee, C. F. & Ostriker, E. C. Annu. Rev. Astron.Astrophys. 45, 565–687 (2007)

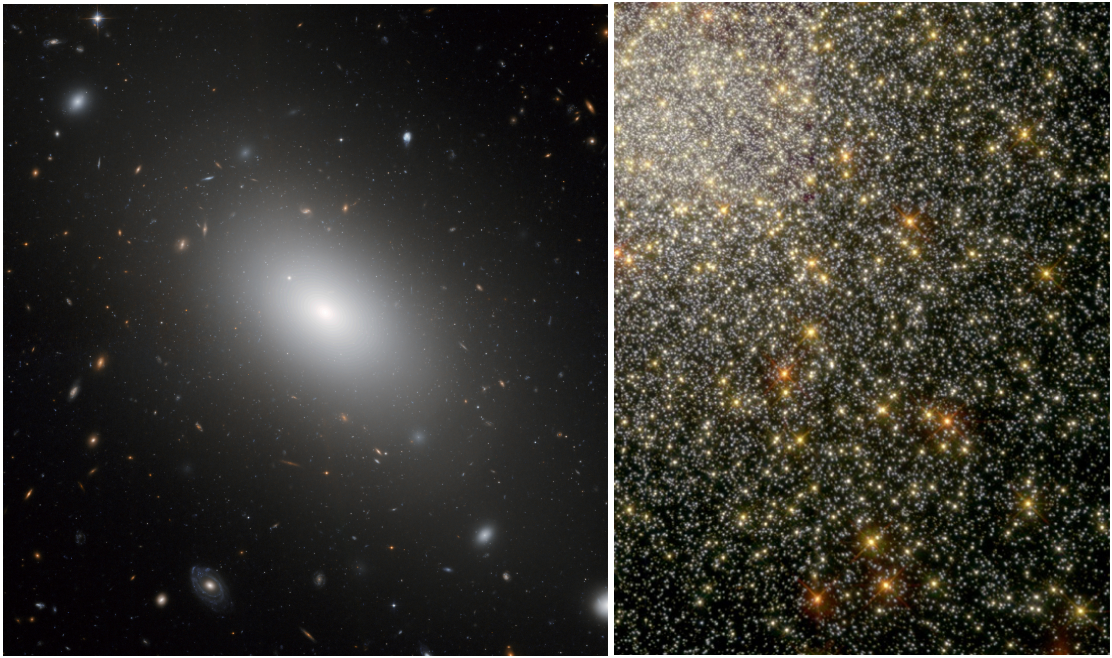

Figure 1: **Viewing galaxies as collections of stars. left**, Cappellari and colleagues' results[2] suggest that giant elliptical galaxies, such as NGC 1132, seen here, have a much higher fraction of low-mass stars than that seen in the local Universe. **right**, Peering into nearby stellar clusters such as 47 Tuc, shown here, reveals stars of different masses; low-mass stars are faint whereas higher-mass stars are bright red or yellow. If 47 Tuc had the same relative distribution of low- to high-mass stars as that suspected for NGC 1132, almost the entire background would be filled with stars.